\documentclass[twocolumn, aps, prb]{revtex4}
\usepackage{graphicx}
\usepackage{amssymb}
\usepackage{amsmath}
\usepackage{color}

\def\Journal #1,#2,#3,#4#5#6#7{#1 {\bf #2}, #3 (#4#5#6#7)}

\def\gsim{\lower -0.3ex \hbox{$>$} \kern -0.75em \lower 0.7ex
\hbox{$\sim$}}
\def\lsim{\lower -0.3ex \hbox{$<$} \kern -0.75em \lower 0.7ex
\hbox{$\sim$}}

\def\Vec#1{{\bf #1}}
\def\GVec#1{\mbox{\boldmath $#1$}}
\def\vare{\varepsilon}

\begin{document}
%%%%%%%%%%%%%%%%%%%%%%%%%%%%%%%%%%%%%%%%%%%%%%%%%%%%%%%%%%%%%%%%%%%%%%%%%%%%%%%
%
%%%%%%%%%%%%%%%%%%%%%%%%%%%%%%%%%%%%%%%%%%%%%%%%%%%%%%%%%%%%%%%%%%%%%%%%%%%%%%%
\title{
Electronic properties of graphene/hexagonal-boron-nitride 
moir\'{e} superlattice 
}
\author{Pilkyung Moon}
\email{pilkyung.moon@nyu.edu}
\affiliation{
New York University Shanghai,
Pudong, Shanghai 200120, China}
\author{Mikito Koshino}
\affiliation{
Department of Physics, Tohoku University, 
Sendai, 980--8578, Japan}

\begin{abstract}
We theoretically investigate the electronic structures of
moir\'{e} superlattices
arising in monolayer / bilayer graphene stacked on 
hexagonal boron nitride (hBN)
in presence and absence of magnetic field.
We develop an effective continuum model
from a microscopic tight-binding lattice Hamiltonian,
and calculate the electronic structures
of graphene-hBN systems with different rotation angles.
Using the effective model,
we explain the characteristic band properties
such as the gap opening 
at the corners of the superlattice Brillouin zone (mini-Dirac
point).
%We generally have a band gap at the 
%corners of the superlattice Brillouin zone (mini-Dirac
%point) due to the inversion symmetry breaking,
%and using the effecive model
%we explain that the gap width is greater in the hole side 
%than in the electron side.
We also investigate the energy spectrum and quantum Hall effect 
of graphene-hBN systems in uniform magnetic field
and demonstrate the evolution 
of the fractal spectrum as a function of the magnetic field.
The spectrum generally splits in the valley degrees of 
freedom ($K$ and $K'$) due to the lack of the inversion symmetry,
and the valley splitting is more significant
in bilayer graphene on hBN than in monolayer graphene on hBN
because of the stronger inversion-symmetry breaking in bilayer.
\end{abstract}
\maketitle

%%%%%%%%%%%%%%%%%%%%%%%%%%%%%%%%%%%%%%%%%%%%%%%%%%%%%%%%%%%%%%%%%%%%%%%%%%%%%%%
%
%%%%%%%%%%%%%%%%%%%%%%%%%%%%%%%%%%%%%%%%%%%%%%%%%%%%%%%%%%%%%%%%%%%%%%%%%%%%%%%
\section{INTRODUCTION}
\label{sec_introduction}

%Since the discovery of two-dimensional
%lattice structure of graphene,
%many of its cousins
%such as hexagonal boron nitride (hBN)\cite{dean2010boron}
%and molybdenum disulphide\cite{radisavljevic2011single}
%have been discovered and widely investigated.
%The two-dimensional nature of these materials
%opens a possibility to make a new kind of artificial structure
%by utilizing the remaining dimension,
%i.e., by layer-by-layer stacking. 
%The structures made in this way
%have plentiful potential in applications
%and are quite versatile in band structure engineering.
Whenever two atomically-thin lattices are stacked
in an incommensurate manner, 
there always arises a superlattice structure
which modulates along the in-plane direction
due to the moir\'{e} interference between different lattice periods.
For example, a bilayer graphene stacked
at an arbitrary angle
(twisted bilayer graphene)\cite{berger2006electronic}
exhibits a periodic variation of the interlayer interaction 
in the form of moir\'{e} pattern,
of which the period can exceed the range of the atomic scale.
The electronic structures of twisted bilayer graphenes
have been intensively investigated, and it is shown that
the material properties, such as the Fermi velocity,
the band energy scale, and optical absorption spectrum,
can be widely tunable with respect to the twist angle.
\cite{lopes2007graphene,trambly2010localization,shallcross2010electronic,morell2010flat,bistritzer2011moirepnas,kindermann2011local,xian2011effects,PhysRevB.86.155449,moon2013opticalabsorption}
Moreover, a huge unit cell of moir\'{e} superlattice provides
an opportunity to investigate the self-similar, 
fractal evolution of the energy spectrum \cite{hofstadter1976energy}
under the simultaneous influences of spatial period
and magnetic field.
\cite{moon2012energy,moon2013opticalproperties,bistritzer2011moireprb}

Recently, the graphene stacked on the hexagonal boron nitride
(hBN) has attracted much attention
as an another moir\'{e} superlattice system.
\cite{dean2010boron,xue2011scanning,yankowitz2012emergence,
ponomarenko2013cloning,hunt2013massive,yu2014hierarchy,yankowitz2014graphene}
Hexagonal boron nitride
is isostructural to graphene,
%and is expected to be free of dangling bonds
%and surface charge traps.
but has boron and nitrogen atoms
at $A$ and $B$ sublattices, respectively,
leading to a finite energy gap in the electronic structure.
\cite{watanabe2004direct,kubota2007deep}
When the graphene is placed on the hBN substrate,
the $1.8\%$ lattice mismatch between graphene and hBN
introduces a superlattice potential
even in a nonrotated stacking.
The transport properties in graphene-hBN systems 
have been investigated experimentally,
and in particular, the fractal electronic structure
was actually observed in magnetic fields.
\cite{dean2013hofstadter,ponomarenko2013cloning,hunt2013massive,yu2014hierarchy}
The electronic structures of graphene-hBN systems
have been studied using
several theoretical approaches.
\cite{sachs2011adhesion,kindermann2012zero,ortix2012graphene,wallbank2013generic,mucha2013heterostructures,PhysRevB.89.075401,bokdam2014band,jung2014ab,San-Jose2014Spontaneous,song2014topological,uchoa2014valley,neek2014graphene,brey2014coherent}
The effective model was derived from the extension from
the twisted bilayer graphene, \cite{kindermann2012zero}
the symmetry based approach \cite{wallbank2013generic,mucha2013heterostructures}
and also from the density functional theory. \cite{jung2014ab}
The energy spectrum in the magnetic field
in the presence of hBN-substrate was also calculated.
\cite{dean2013hofstadter,ponomarenko2013cloning,hunt2013massive,yu2014hierarchy}

In this paper,
we theoretically investigate the electronic structures of
moir\'{e} superlattices
arising in monolayer and bilayer graphene stacked on hBN layer
with and without magnetic field.
We develop an effective continuum model
starting from a microscopic tight-binding lattice Hamiltonian,
and calculate the electronic structures
of graphene-hBN systems with several different rotation angles.
The model is expressed in terms of a few parameters, 
which are analytically extracted from the microscopic parameters
in the given tight-binding model.
%such as the strength of the interlayer coupling
%as well as the on-site energies of B and N.
We verify the validity of the effective model
by demonstrating that the calculated band structure 
agrees with that of the original tight-binding model.
In the band structure calculation,
we find that we generally have a band gap at the 
zone corners of the superlattice Brillouin zone (so-called mini-Dirac
point) due to the inversion symmetry breaking,
and the gap width is shown to 
be greater in the hole side than in the electron side.
We analytically explain the origin of
the electron-hole asymmetric gap opening 
in terms of the matrix elements of the effective model.

We then study the energy spectrum and quantum Hall effect 
of graphene-hBN systems in uniform magnetic fields,
and demonstrate the evolution 
of the fractal spectrum as a function of the magnetic field.
We find that the spectrum generally splits in the valley degrees of 
freedom ($K$ and $K'$) due to the lack of the inversion symmetry.
The valley splitting is more significant
in bilayer graphene on hBN than in monolayer graphene on hBN,
because the inversion symmetry is severely broken
in the bilayer case, where only a single layer out of two graphene layers
feels the effective potential from hBN.

The paper is organized as follow.
In Sec.\ \ref{theoretical_methods},
we derive an effective continuum model
for graphene on hBN structures
from a tight-binding Hamiltonian.
On the basis of the effective
and tight-binding models,
we study the band structures
of both monolayer and bilayer graphene
on hBN in Sec.\ \ref{band_structure}.
In Sec.\ \ref{spectrum_in_magnetic_field},
we investigate the fractal energy spectrum
and the quantum Hall effect of electrons under magnetic field.
Finally, conclusions are given in Sec.\ \ref{conclusion}.

%The difference in the origin of the gaps
%at Dirac point of monolayer and bilayer graphene
%on hBN will be investigated.
%Moreover, the origin of the small but finite
%difference between the band structures
%of two different configurations
%in AB-stacked bilayer graphene on hBN
%will be explained analytically.
%The present effective model shares the same theoretical basis
%with the model developed for the graphene-graphene bilayer system, 
%\cite{moon2013opticalabsorption} and therefore
%it provides a unified picture which covers these different moir\'{e} systems.

\section{Theoretical methods}
\label{theoretical_methods}

\subsection{Atomic structure and moir\'{e} lattice vectors}
\label{subsection:Atomic structure and Brillouin zone}

We consider a bilayer system composed of graphene and hBN.
Graphene is a two-dimensional honeycomb lattice of carbon atoms,
of which the unit cell includes $A$ and $B$ sublattices.
The hBN is a similar honeycomb lattice but composed of
nitride atom on $A$-site and boron atom on $B$-site.
The lattice constant (i.e., the distance between
the nearest $A$-sites) of hBN
is given by $a_{\rm hBN} \approx 0.2504\,\mathrm{nm}$,\cite{liu2003structural}
which is slightly larger than
$a \approx 0.246\,\mathrm{nm}$ for graphene.
We assume that the interlayer distance between graphene and hBN
is constant at $d_{\rm G-hBN} =0.322\,\mathrm{nm}$.\cite{giovannetti2007substrate}

We define the stacking geometry of the 
graphene-hBN bilayer system by starting from 
a nonrotated arrangement,
where a $B$-site of graphene and a $B$-site of hBN share 
the same in-plane position $(x,y)=0$, and 
the $A$-$B$ bonds are parallel to each other.
We then  rotate the hBN with respect to graphene
by an arbitrary angle $\theta$ around the origin.
We define $\Vec{a}_1 = a(1,0)$ and $\Vec{a}_2 = a(1/2,\sqrt{3}/2)$ 
as the lattice vectors of graphene.
The primitive lattice vectors of hBN become
\begin{equation}
 \tilde{\Vec{a}}_i = M R \,\, \Vec{a}_i \quad (i=1,2),
\end{equation}
where $R$ is the rotation matrix by $\theta$,
and $M = (1+\vare) \Vec{1}$ 
represents the isotropic expansion by the factor 
$1+\vare = a_{\rm hBN}/a \approx 1.018$.
We define the reciprocal lattice vectors 
$\Vec{a}^*_i$ and $\tilde{\Vec{a}}^*_i$ for graphene and hBN, respectively,
so as to satisfy
$\Vec{a}_i\cdot\Vec{a}^*_j = \tilde{\Vec{a}}_i\cdot\tilde{\Vec{a}}^*_j = 2\pi\delta_{ij}$.

\begin{figure}
\begin{center}
%\leavevmode\includegraphics[width=0.9\hsize]{fig_unit_cell.eps}
\leavevmode\includegraphics[width=0.9\hsize]{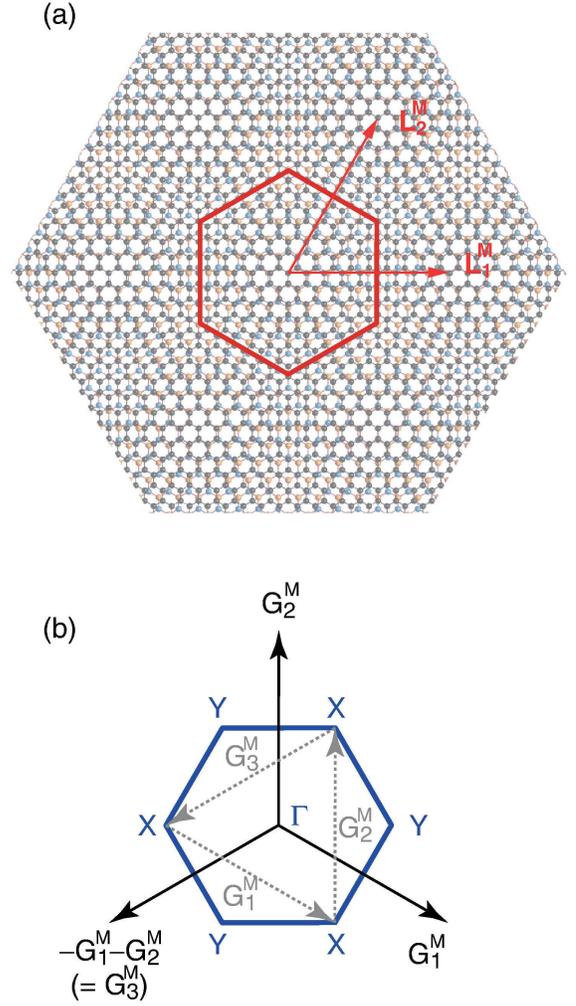}
\end{center}
\caption{
(a) Graphene-hBN moir\'{e} superlattice
with $\theta =0^\circ$
and an exaggerated lattice constant ratio $a_{\rm hBN}/a = 10/9$.
Unit cell is indicated by a hexagon.
(b) Superlattice Brillouin zone 
with the reciprocal lattice vectors
spanned by $\Vec{G}^{\rm M}_i$.
}
\label{fig_unit_cell}
\end{figure}

The mismatch of the lattice periods of 
graphene and hBN gives rise to the moir\'{e} interference pattern.
An atom on hBN located at position $\Vec{r}$ has its counterpart 
on graphene at $R^{-1}M^{-1} \,\, \Vec{r}$. 
The displacement vector between two sites (from graphene to hBN) 
is
\begin{equation}
 \GVec{\delta}(\Vec{r}) =
 (\Vec{1}-R^{-1}M^{-1})\Vec{r}.
\label{eq_delta_of_r}
\end{equation}
When $\GVec{\delta}(\Vec{r})$ coincides with a lattice vector
of graphene, then graphene and hBN share the same phase of the
lattice periodicity (i.e., the corresponding positions of their hexagonal unit cells) 
at the position $\Vec{r}$, in the same way as in the origin.
Therefore, the primitive lattice vector of the moir\'{e}
superlattice $\Vec{L}_i^{\rm M}$ is obtained from the condition
$\delta(\Vec{L}_i^{\rm M}) = \Vec{a}_i$, which leads to
\begin{equation}
  \Vec{L}_i^{\rm M} = (\Vec{1}-R^{-1}M^{-1})^{-1} \Vec{a}_i\quad (i=1,2).
\end{equation}
The corresponding moir\'{e} reciprocal lattice vectors 
satisfying $\Vec{G}^{\rm M}_i\cdot\Vec{L}_j^{\rm M} = 2\pi\delta_{ij}$
are written as
\begin{equation}
  \Vec{G}_i^{\rm M} = (\Vec{1}-M^{-1}R) \Vec{a}^*_i\quad (i=1,2),
\end{equation}
where we used $R^\dagger = R^{-1}$ and $M^\dagger = M$.
The moir\'{e} lattice period 
$ L^{\rm M} = |\Vec{L}_1^{\rm M}| = |\Vec{L}_2^{\rm M}|$
is \cite{yankowitz2012emergence}
\begin{equation}
 L^{\rm M}  
= \frac{1+\vare}{\sqrt{\vare^2+2(1+\vare)(1-\cos\theta)}}\, a,
\end{equation}
and the angle from $\Vec{a}_i$ to $\Vec{L}_i^{\rm M}$ is 
\begin{equation}
 \phi = {\rm arctan}\left(
\frac{-\sin\theta}{1+\vare-\cos\theta}
\right).
\label{eq_angle_phi}
\end{equation}
When $\theta=0^\circ$, we have $L^{\rm M} = 13.8\,\mathrm{nm}$.
Figure \ref{fig_unit_cell}(a)
shows the atomic structure and unit cell
of graphene-hBN moir\'{e} with $\theta =0^\circ$
and an exaggerated lattice constant ratio $a_{\rm hBN}/a = 10/9$.
Figure \ref{fig_unit_cell}(b) is the superlattice Brillouin zone
spanned by $\Vec{G}^{\rm M}_i$.

\subsection{Tight-binding model}
\label{subsection:tight-binding_model}

We consider the tight-binding model for $p_z$ atomic orbitals.
The Hamiltonian is written as
\begin{eqnarray}
 H = -\sum_{i,j}
t(\Vec{R}_i - \Vec{R}_j)
|\Vec{R}_i\rangle\langle\Vec{R}_j| 
+ \sum_{i}V(\Vec{R}_i) |\Vec{R}_i\rangle\langle\Vec{R}_i|,
\label{eq_TB_hamiltonian}
\end{eqnarray}
where $\Vec{R}_i$ and $|\Vec{R}_i\rangle$ 
represent the lattice point and the atomic state at site $i$, respectively,
$V(\Vec{R}_i)$ is the on-site potential at site $i$,
and $t(\Vec{R}_i - \Vec{R}_j)$ is
the transfer integral between the sites $i$ and $j$. 
We assume $V_{\rm C} = 0$ for carbon atom, and
\begin{equation}
V_{\rm B} = 3.34\,{\rm eV},
\quad
V_{\rm N} = -1.40\,{\rm eV},
\end{equation}
for boron and nitride atoms, respectively.\cite{slawinska2010energy}

For the transfer integral, we simply adopt the common
Slater-Koster-type function for any combinations of atomic species,
\cite{nakanishi2001conductance,uryu2004electronic,trambly2010localization,slater1954simplified}
\begin{eqnarray}
 && -t(\Vec{R}) = 
V_{pp\pi}\left[1-\left(\frac{\Vec{R}\cdot\Vec{e}_z}{R}\right)^2\right]
+ V_{pp\sigma}\left(\frac{\Vec{R}\cdot\Vec{e}_z}{R}\right)^2,
\nonumber \\
&& V_{pp\pi} =  V_{pp\pi}^0 % -\tilde\gamma_0
\exp \left(- \frac{R-a_0}{r_0}\right),
\nonumber \\
&& V_{pp\sigma} =  V_{pp\sigma}^0 %\tilde\gamma_1
 \exp \left(- \frac{R-d_0}{r_0}\right),
\label{eq_transfer_integral}
\end{eqnarray}
Here $\Vec{e}_z$ is the unit vector
perpendicular to the graphene plane,
$a_0 = a/\sqrt{3} \approx 0.142\,\mathrm{nm}$ is the distance of
neighboring $A$ and $B$ sites on graphene,
and $d_0 \approx 0.335\,\mathrm{nm}$
is the interlayer spacing of graphene.
$V_{pp\pi}^0$ is the transfer integral between 
the nearest-neighbor atoms of monolayer graphene
and $V_{pp\sigma}^0$ is that
between vertically located atoms on the neighboring layers. 
We take $V_{pp\pi}^0 \approx -2.7\,\mathrm{eV}$,
$V_{pp\sigma}^0 \approx 0.48\,\mathrm{eV}$, to
fit the dispersions of monolayer graphene and $AB$-stacked bilayer 
graphene.\cite{trambly2010localization}
$r_0$ is the decay length of the transfer integral,
and is chosen as $0.184 a$ so that 
the next nearest intralayer coupling becomes $0.1 V_{pp\pi}^0$.
\cite{uryu2004electronic,trambly2010localization} 
%The transfer integral for $d > 4 a_0$ is exponentially small 
%and can be safely neglected.
%If we adopt above formula to graphene-hBN system, 
%the interlayer hopping becomes slightly bigger
%than bilayer graphene because we replace $d$ with $d_{\rm G-BN}$
%in Eq.\ (\ref{eq_transfer_integral}).

In the tight-binding band calculation,
the lattice structure of a graphene-hBN composite system
must have a finite unit cell,
and for this purpose we take $\theta = 0$
and rationalize the relative lattice period
$a_{\rm hBN}/a\approx 1.018$ to 56/55.
%where the moir\'{e} superlattice period becomes $L^{\rm M}=56a$.
We do not need the lattice rationalization
in the continuum model argued in the next section, 
where the atomic period $a$ is smeared out and 
the Hamiltonian is governed only by $L^{\rm M}$.

\begin{figure}
\begin{center}
\leavevmode\includegraphics[width=0.65\hsize]{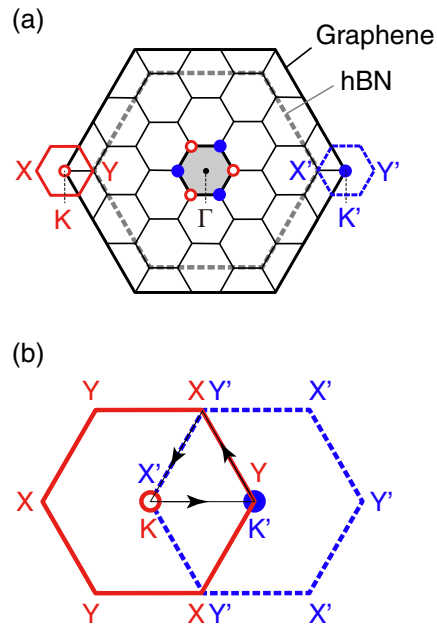}
\end{center}
\caption{
(a) Brillouin zone (BZ) folding
in graphene-hBN moir\'{e} superlattice,
where $a_{\rm hBN}/a$ is taken as 5/4 
for the illustrative purpose.
(b) Relative positions of
the BZs centered at $K$ and $K'$ valleys
in the common BZ.
}
\label{fig_bz}
\end{figure}

Figure \ref{fig_bz}(a) 
illustrates the Brillouin zone (BZ) folding
where $a_{\rm hBN}/a$ is taken as 5/4 (instead of 56/55) 
for the illustrative purpose.
The solid large hexagon is the graphene's BZ
spanned by $\Vec{a}^*_i$,
%the dashed large hexagon is the hBN's BZ
%spanned by $\tilde{\Vec{a}}^*_i$,
and the small hexagon is the reduced BZ
spanned by $\Vec{G}^{\rm M}_i$.
In the tight-binding model, 
$K$ and $K'$ are inseparable and all the energy bands 
are folded in the common BZ.
In the continuum model, on the other hand, 
$K$ and $K'$ valleys are treated independently,
and the energy bands can be separately plotted 
in the BZ centered at $K$ (red, solid)
and that centered at $K'$ (blue, dashed).
Figure \ref{fig_bz}(b) shows the relation between
the separate BZs in the original common BZ.
\footnote[1]{The relative position of $K$ and $K'$
in the folded BZ actually depends on the 
rounded value of $L^{\rm M}/a$ in the modulo 3,
and it is equivalent in the exaggerated $a_{\rm hBN}/a=5/4$ and
in the original $56/55$.
}

\subsection{Effective continuum model}
\label{subsection:effective_continuum_model}

%In the following, we assume
%the twist angle $\theta$ is so small that the moir\'{e} superlattice period 
%$L^{\rm M}$ is much larger than the lattice constant $a$.

When the rotation angle $\theta$ is small
and the moir\'{e} superlattice period $L^{\rm M}$
is much larger than the lattice constant $a$,
the interaction between the two graphene layers
is dominated by long-wavelength components,
allowing one to treat the problem in the 
effective continuum model.
In the literature,
the continuum approach has been introduced 
for the twisted graphene-graphene bilayer
\cite{lopes2007graphene,bistritzer2011moirepnas,kindermann2011local,PhysRevB.86.155449,moon2013opticalabsorption}
and also for the graphene-hBN system.\cite{kindermann2012zero,wallbank2013generic,jung2014ab}
Here we derive an effective continuum model
starting from the microscopic tight-binding Hamiltonian
using the approach developed for the twisted graphene
bilayer.\cite{moon2013opticalabsorption}
The effective Hamiltonian is expressed 
in terms of a few parameters, 
which are directly extracted from the microscopic parameters
in the given tight-binding model.

The low-energy spectrum of graphene is dominated by
the states near $K$ and $K'$ points,
and the effective Hamiltonian is approximated by 
the effective Dirac cones centered at those points.
\cite{mcclure1956diamagnetism,divincenzo1984self,semenoff1984condensed,shon1998quantum,ando2005theory}
In the present case, the $K$ points of graphene are located at
$\Vec{K}_{\xi} = -\xi (2\Vec{a}^*_1+\Vec{a}^*_2)/3$ while 
$\xi=\pm 1$ for $K$ and $K'$, respectively.
The Hamiltonian of monolayer graphene 
near $\Vec{K}_\xi$ is written as
\begin{eqnarray}
 && H_{\rm G} \approx -\hbar v {\Vec{k}} %({\Vec{k}}-\Vec{K}_\xi)
\cdot \GVec{\sigma}_\xi,
\label{eq_H_MLG}
\end{eqnarray}
where $\Vec{k}$ is the relative wave number
measured from $\Vec{K}_\xi$ point,
and $\GVec{\sigma}_\xi = (\xi \sigma_x, \sigma_y)$
with Pauli matrices $\sigma_x$ and $\sigma_y$.
The parameter $v$ is the band velocity of the Dirac cone,
which is given in the present tight-binding parametrization as
$v \approx
(\sqrt{3}a/2\hbar) V_{pp\pi}^0(1-2e^{-a_0/r_0})
\approx 0.80\times 10^6$ m/s.
\cite{moon2013opticalabsorption}

$K$ valleys of hBN are given by $\tilde{\Vec{K}}_{\xi} = -\xi
(2\tilde{\Vec{a}}^*_1+\tilde{\Vec{a}}^*_2)/3$.
The effective Hamiltonian of hBN monolayer includes
a similar kinetic term linear to the relative wave number from 
$\tilde{\Vec{K}}_\xi$,
plus the on-site potential term $V_{\rm N}$ and $V_{\rm B}$.
This gives a massive Dirac cone separated by an energy gap
$V_{\rm B}-V_{\rm N}$
with a quadratic dispersion centered at $\tilde{\Vec{K}}_\xi$.
Here we adopt an approximation
in which we completely neglect the dispersion of hBN 
by dropping $\Vec{k}$,\cite{kindermann2012zero} i.e.,
\begin{eqnarray}
&& H_{\rm hBN} \approx 
% -\hbar v_{\rm hBN} [R^{-1}M({\Vec{k}}-\tilde{\Vec{K}}_\xi)]
%\cdot (\xi \sigma_x, \sigma_y)
%\nonumber\\
%&& \qquad\qquad\qquad
%+
\begin{pmatrix}
V_{\rm N} & 0 \\
0 & V_{\rm B}
\end{pmatrix}.
\label{eq_H_hBN}
\end{eqnarray}
This is justified when $\theta$ is small,
because ${\Vec{K}}_{\xi}$ and $\tilde{\Vec{K}}_{\xi}$ are 
then close to each other,
and the graphene's electronic states near ${\Vec{K}}_{\xi}$
are coupled only with the hBN's states near $\tilde{\Vec{K}}_{\xi}$
by the long-range interlayer coupling.

The interlayer coupling term between graphene and hBN
can be derived in a similar manner to the twisted bilayer 
graphene.\cite{moon2013opticalabsorption}
When $\theta$ is small, the local lattice structure is 
approximately viewed as a pair of identical honeycomb lattices 
shifted by a displacement vector $\GVec{\delta}$
with no rotation.
$\GVec{\delta}$ slowly depends on the position
$\Vec{r}$ in accordance with Eq.\ (\ref{eq_delta_of_r}).
The interlayer coupling term for the nonrotated honeycomb bilayer 
with a constant $\GVec{\delta}$ can be derived 
from a tight-binding model in a straightforward manner,
which is described in the Appendix \ref{sec_app_a}.
By replacing constant $\GVec{\delta}$ with $\GVec{\delta}(\Vec{r})$,
we obtain the interlayer Hamiltonian for the moir\'{e} system.
As a result, the effective Hamiltonian of the graphene-hBN system 
near the $\Vec{K}_\xi$ point
is written as
\begin{eqnarray}
 {\cal H}_{\rm G-hBN} = 
\begin{pmatrix}
H_{\rm G} & U^\dagger \\
U & H_{\rm hBN}
\end{pmatrix},
\label{eq_H_G-hBN}
\end{eqnarray}
with
\begin{eqnarray}
 && U = 
\begin{pmatrix}
U_{A_2 A_1} & U_{A_2 B_1}
\\
U_{B_2 A_1} & U_{B_2 B_1}
\end{pmatrix}
=
u_0
\Biggl[
\begin{pmatrix}
1 & 1
\\
1 & 1
\end{pmatrix}
+
\nonumber\\
&&
\quad
\begin{pmatrix}
1 & \omega^{-\xi}
\\
\omega^{\xi} & 1
\end{pmatrix}
e^{i\xi\Vec{G}^{\rm M}_1\cdot\Vec{r}}
+
\begin{pmatrix}
1 & \omega^{\xi}
\\
\omega^{-\xi} & 1
\end{pmatrix}
e^{i\xi(\Vec{G}^{\rm M}_1+\Vec{G}^{\rm M}_2)\cdot\Vec{r}}
\Biggr],
\nonumber\\
\label{eq_U}
\end{eqnarray}
%\begin{eqnarray}
%&& U =
%\begin{pmatrix}
%U_{A_2 A_1} & U_{A_2 B_1}
%\\
%U_{B_2 A_1} & U_{B_2 B_1}
%\end{pmatrix},
%\nonumber\\
%&& U_{A_2 A_1} = U_{B_2 B_1} 
%= u_0 [
%1
% +e^{i\xi\Vec{G}^{\rm M}_1\cdot\Vec{r}}
%+ e^{i\xi(\Vec{G}^{\rm M}_1+\Vec{G}^{\rm M}_2)\cdot\Vec{r}}
%],
%\nonumber\\
%&& U_{A_2 B_1} 
%= u_0 [
%1
%+ \omega^{-\xi}e^{i\xi\Vec{G}^{\rm M}_1\cdot\Vec{r}}
%+ \omega^{\xi} e^{i\xi(\Vec{G}^{\rm M}_1+\Vec{G}^{\rm M}_2)\cdot\Vec{r}}
%],
%\nonumber\\
%&& U_{B_2 A_1}
%= u_0 [
%1
%+ \omega^{\xi}e^{i\xi\Vec{G}^{\rm M}_1\cdot\Vec{r}}
%+ \omega^{-\xi}e^{i\xi(\Vec{G}^{\rm M}_1+\Vec{G}^{\rm M}_2)\cdot\Vec{r}}
%],
%\label{eq_U}
%\end{eqnarray}
and $\omega = \exp(2\pi i/3)$.
Here the $4\times 4$ matrix is written for the basis of $\{A_1,B_1,A_2,B_2 \}$,
with $A_1,B_1$ for graphene, $A_2,B_2$ for hBN.
The only parameter $u_0$ is defined by 
the in-plane Fourier transform of the transfer integral $t(\Vec{R})$,
\begin{eqnarray}
 u_0 = 
\frac{1}{S} \int t(\Vec{R}+ \Vec{d}_z) 
e^{-i \Vec{\Vec{K}_\xi}\cdot \Vec{R}} d\Vec{R},
\label{eq_u0}
\end{eqnarray}
where $S = |\Vec{a}_1\times\Vec{a}_2|$ is the unit cell area, 
$\Vec{d}_z = d_{\rm G-hBN}\,\Vec{e}_z$ is the 
perpendicular displacement between graphene and hBN,
and the integral in $\Vec{R}$
is taken over an infinite two-dimensional space.
The $u_0$ does not depend on $\xi$, 
and we have $u_0 \approx 0.152\,\mathrm{eV}$
in the present tight-binding parameters.

%The parameter $u_0$ is related to the hopping integral via
%\begin{eqnarray}
%u_0 = \frac{1}{3}t(0,d) - t(a,d) + 2t(\sqrt{3}a,d) -\cdots,
%\end{eqnarray}
%where $t(x,z) = t(x \Vec{e}_x + z \Vec{e}_z)$.
%By using the tight-binding parameters
%and the interlayer spacing $d$, we have $u_0 \approx 0.152$eV. 

Since the energy band of hBN is gapped,
the low-energy spectrum near $E\approx 0 $
is dominated by graphene's electronic states.
Then the effective Hamiltonian is even reduced to
a $2 \times 2$ form by eliminating the hBN bases
by the second order perturbation. 
The results is,
\begin{eqnarray}
 {\cal H}^{\rm (red)}_{\rm G-hBN} &=& 
H_{\rm G} + U^\dagger (-H_{\rm hBN})^{-1} U
\nonumber\\
&\equiv&
H_{\rm G} +  V_{\rm hBN},
%=\begin{pmatrix}
%H_{\rm G} & U^\dagger \\
%U & H_{\rm hBN}
%\end{pmatrix},
\label{eq_H_G-hBN_eff}
\end{eqnarray}
where the additional term $V_{\rm hBN}$
is explicitly written as
\begin{eqnarray}
&& V_{\rm hBN} = 
V_0
\begin{pmatrix}
1 & 0
\\
0 & 1
\end{pmatrix}
\nonumber\\
&&
+
\Biggl\{
V_1 e^{i\xi\psi}
\Biggl[
\begin{pmatrix}
1 & \omega^{-\xi}
\\
1 & \omega^{-\xi}
\end{pmatrix}
e^{i\xi\Vec{G}^{\rm M}_1\cdot\Vec{r}}
+
\begin{pmatrix}
1 & \omega^{\xi}
\\
\omega^{\xi} & \omega^{-\xi}
\end{pmatrix}
e^{i\xi\Vec{G}^{\rm M}_2\cdot\Vec{r}}
\nonumber\\
&&
\qquad\quad
+
\begin{pmatrix}
1 & 1
\\
\omega^{-\xi} & \omega^{-\xi}
\end{pmatrix}
e^{-i\xi(\Vec{G}^{\rm M}_1+\Vec{G}^{\rm M}_2)\cdot\Vec{r}}
\Biggr]
\,\,\,+ {\rm h.c.} \Biggr\},
\label{eq_VhBN}
\end{eqnarray}
and 
\begin{eqnarray}
 && V_0 = -3 u_0^2 
\left(
\frac{1}{V_{\rm N}}
+ \frac{1}{V_{\rm B}}
\right),
\\
 && V_1 e^{i\psi} = - u_0^2 
\left(
\frac{1}{V_{\rm N}}
+ \omega \frac{1}{V_{\rm B}}
\right).
\label{eq_v1}
\end{eqnarray}
%In Eq.\ (\ref{eq_v1}), 
%$V_1$ represents the amplitude and 
%$\psi$ a phase factor.
In the present parameters, we have
$V_0 \approx 0.0289\,\mathrm{eV}$,
$V_1 \approx 0.0210\,\mathrm{eV}$,
and $\psi \approx -0.29$(rad).

The effective term $V_{\rm hBN}$ can be generally divided into
the scalar potential $V^{\rm eff}$,
the vector potential $\Vec{A}^{\rm eff}$,
and the Dirac mass term 
$M^{\rm eff}$ as \cite{kindermann2012zero,wallbank2013generic} 
\begin{eqnarray}
 V_{\rm hBN}  = 
V^{\rm eff}(\Vec{r}) 
+ M^{\rm eff}(\Vec{r})  \sigma_z
+ ev\Vec{A}^{\rm eff}(\Vec{r})  \cdot \GVec{\sigma}_\xi.
\label{eq_VhBN_decompose}
\end{eqnarray}
For the present $V_{\rm hBN}$ of Eq.\ (\ref{eq_VhBN}), we have
\begin{eqnarray}
&& 
V^{\rm eff}(\Vec{r}) =
V_0 - V_1 \sum_{l=1}^3 
\cos \alpha_l(\Vec{r})
\nonumber\\
&& M^{\rm eff}(\Vec{r}) =
\sqrt{3} V_1 \sum_{l=1}^3 
\sin \alpha_l(\Vec{r})
\nonumber\\
&&
ev \Vec{A}^{\rm eff} (\Vec{r}) =
2 \xi V_1
 \sum_{l=1}^3 
\begin{pmatrix}
 \cos\left[2\pi(l+1)/3\right]\\
 \sin\left[2\pi(l+1)/3\right]
\end{pmatrix}
\cos \alpha_l(\Vec{r})
\nonumber\\
\label{eq_effective_terms}
\end{eqnarray}
with
\begin{equation}
\alpha_l (\Vec{r})
= \Vec{G}_l^{\rm M}\cdot \Vec{r} + \psi + \frac{2\pi}{3}, 
\end{equation}
where we defined  
$\Vec{G}_3^{\rm M} = -\Vec{G}_1^{\rm M}-\Vec{G}_2^{\rm M}$
so that the vectors $\Vec{G}_1^{\rm M}$,
$\Vec{G}_2^{\rm M}$ and $\Vec{G}_3^{\rm M}$
are pointing to the trigonal symmetric directions.
The effective vector potential $\Vec{A}^{\rm eff}$
gives the effective magnetic field,
\begin{eqnarray}
&& 
B^{\rm eff}(\Vec{r}) 
= \nabla \times \Vec{A}^{\rm eff}
= \xi B_0 
\sum_{l=1}^3 
\cos \alpha_l (\Vec{r}),
\end{eqnarray}
where
\begin{equation}
 B_0 = \frac{2 V_1 G^{\rm M}}{ev} \cos\phi.
\end{equation}
Here $G^{\rm M} \equiv |\Vec{G}_l^{\rm M}| = (4\pi/\sqrt{3})/L^{\rm M}$,
and $\phi$ is defined by Eq.\ (\ref{eq_angle_phi}).
The effective magnetic field is opposite between the different valleys
$\xi = \pm 1$ so as to satisfy the time-reversal symmetry.
In the present model, the magnitude of the effective magnetic field
is $B_0 \sim 0.022$T at $\theta =0^\circ$.

% inversion symmetry

The lattice structure of the graphene-hBN hybrid system is not invariant
in the spatial inversion, and accordingly, 
the effective Hamiltonian Eq.\ (\ref{eq_H_G-hBN_eff})
lacks the inversion symmetry as shown in the following.
The spatial inversion 
changes  $\Vec{r}$ to $-\Vec{r}$,
and at the same time it swaps $A$ and $B$ sublattices 
and $K$ and $K'$ valleys.
If the system has the inversion symmetry, then
the Hamiltonian $H^{(\xi)}(\Vec{k},\Vec{r})$ yields to
\begin{equation}
 H^{(-\xi)}(\Vec{k},\Vec{r}) = 
\sigma_x [ H^{(\xi)}(-\Vec{k},-\Vec{r}) ]\sigma_x.
\label{eq_inv_sym}
\end{equation}
The Hamiltonian of pristine graphene, Eq.\ (\ref{eq_H_MLG}), 
satisfies this condition.
For the effective potential terms in Eq.\ (\ref{eq_VhBN_decompose}),
the condition Eq.\ (\ref{eq_inv_sym}) is rewritten as
$V^{\rm eff}(-\Vec{r}) = V^{\rm eff}(\Vec{r})$,
$M^{\rm eff}(-\Vec{r}) = -M^{\rm eff}(\Vec{r})$, and
$\Vec{A}^{\rm eff}(-\Vec{r}) = \Vec{A}^{\rm eff}(\Vec{r})$.
The effective terms in Eq.\ (\ref{eq_effective_terms})
meet these conditions only when $\psi + 2\pi/3 =n\pi$ ($n$: integer),
while it is not the case in the present model
($\psi + 2\pi/3 \approx 0.57\pi$).

% AB-bilayer + hBN

We can also consider $AB$-stacked bilayer graphene + hBN monolayer system
using the same approach.
Here we number the layer 1, 2 for graphene bilayer
where the layer 1 is faced to the hBN layer,
and assume that the two graphene layers are stacked
so that $B_1$-site and $A_2$-site
are vertically located as shown in Fig.\ \ref{fig_band_bi_diff}(a).
After eliminating the hBN bases in a similar manner,
the effective Hamiltonian in the basis of $\{A_1,B_1,A_2,B_2 \}$ 
is written as
\begin{eqnarray}
 {\cal H}_{\rm BLG-hBN} = 
\begin{pmatrix}
H_{\rm G} + V_{\rm hBN} & U_{\rm BLG}^\dagger \\
U_{\rm BLG} & H_{\rm G} 
\end{pmatrix},
\label{eq_H_BLG-hBN}
\end{eqnarray}
where $U_{\rm BLG}$ is
the interlayer coupling between $AB$-stacked graphenes,\cite{mccann2006landau}
\begin{eqnarray}
 U_{\rm BLG} = 
\begin{pmatrix}
0  & \gamma_1 \\
-\hbar v_3(\xi k_x -i k_y) & 0
%-\hbar v_3[\xi(k_x-K^{(\xi)}_x) -i (k_y-K^{(\xi)}_y)] & 0
\end{pmatrix}.
\end{eqnarray}
The parameter $\gamma_1$ represents the band splitting
and $v_3$ describes the trigonal warping.\cite{mccann2006landau}
In the present tight-binding parameters, we have $\gamma_1 = 0.34\,\mathrm{eV}$ and
$v_3 = 0.051\times 10^6$m/s.

We have another possibility of graphene $AB$-stacking 
in which $A_1$-site and $B_2$-site
are vertically located,
as shown in Fig.\ \ref{fig_band_bi_diff}(b).
This is just 180$^\circ$ in-plane rotation of the previous 
$B_1$-$A_2$ stacking, but they are not equivalent
when hBN is added to the third layer,
since
neither graphene bilayer nor hBN are invariant
in 180$^\circ$ rotation.
The effective Hamiltonian for the second case is
obtained by interchanging the off-diagonal blocks
in Eq.\ (\ref{eq_H_BLG-hBN}).
We find the energy spectra of the two models 
make no qualitative difference, although they are not identical.
In the following calculation, 
we concentrate on the case of Eq.\ (\ref{eq_H_BLG-hBN}),
while the energy spectrum for the second case 
is argued in Appendix \ref{subsection:ab_bilayer_and_ba_bilayer_with_hBN}.

\section{Band structure}
\label{band_structure}

\begin{figure*}
\begin{center}
%\leavevmode\includegraphics[width=0.9\hsize]{fig_band_mono.eps}
\leavevmode\includegraphics[width=0.9\hsize]{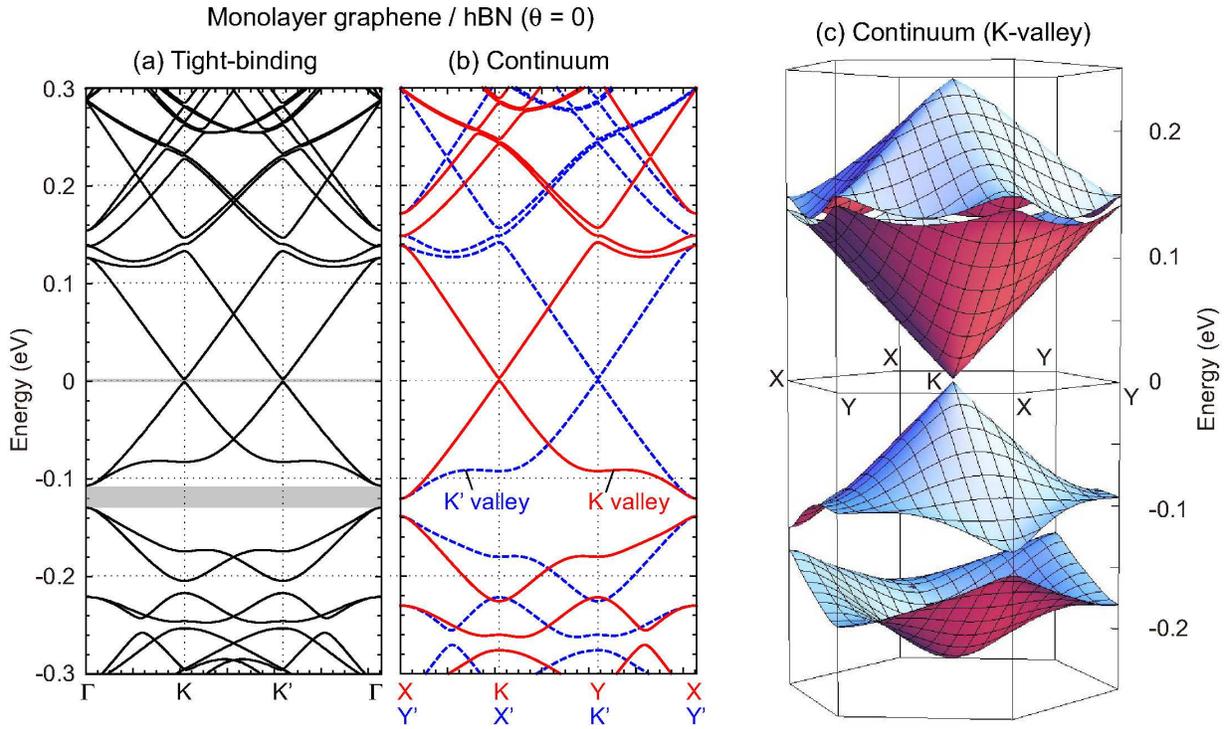}
\end{center}
\caption{
Band structures
of monolayer graphene / hBN system with $\theta=0^\circ$
calculated by (a) the tight-binding model 
and (b) the effective continuum model,
on the $k$-space path shown in Fig.\ \ref{fig_bz}(b).
(c) Three-dimensional plot
of the first and second electron and hole bands of $K$-valley,
calculated by the continuum model.
}
\label{fig_band_mono}
\end{figure*}

First, we calculate the band structure of monolayer graphene-hBN 
system at $\theta=0^\circ$,
both in the tight-binding model and in the effective continuum model.
Figures \ref{fig_band_mono}(a) and \ref{fig_band_mono}(b) compare
the energy band structure
calculated by the tight-binding model (with $a_{\rm hBN}/a=56/55$)
and the effective continuum model, respectively,
on the $k$-space path shown in Fig.\ \ref{fig_bz}(b).
Here and in the following, the origin of the energy axis
is reset to the charge neutral point.
We see that the agreement between the two models 
is almost complete, showing that 
the effective continuum model describes
the detail of the low-energy spectrum quantitatively well.
Figure \ref{fig_band_mono}(c) is the three-dimensional plot
of the first and second electron and hole bands of $K$-valley,
calculated by the continuum model.

In the spectrum, we see a band splitting between the first 
and the second electron (hole) bands,
due to the band anticrossing at the Brillouin zone boundary.
%The energy position of the band edge
%is estimated as $\pm (4\pi/3)(\hbar v/L^{\rm M}) \approx \pm 0.16\,\mathrm{eV}$
%in the limit of zero interlayer coupling,
%although the actual energy is a little lower than this
%due to the level repulsion.
The splitting is fairly large in the hole side 
leading to an actual spectral gap from $E=-0.14\,\mathrm{eV}$
to $-0.12\,\mathrm{eV}$,
while it is much narrower in the electron side.
This feature is consistent with
the experiments, showing that the hole side exhibits 
a stronger resistance peak than the electron side.
\cite{dean2013hofstadter,yankowitz2012emergence,yu2014hierarchy}
At the central Dirac point, 
there is a tiny energy gap about $2\,\mathrm{meV}$,
which is proportional to the third order 
to the interlayer coupling $u_0$. \cite{kindermann2012zero}  
%$u_0^3/(2\pi \hbar v/L^{\rm M})^2$,
It should be mentioned that
the recent experiments \cite{hunt2013massive,woods2014commensurate} 
reported that a much larger bandgap opens at the central Dirac point 
in graphene/hBN systems with small twist angles.
There are several theoretical approaches
to explain the origin of the band gap in terms of the strain effect
\cite{jung2014ab,bokdam2014band,San-Jose2014Spontaneous}
and many-body interaction \cite{bokdam2014band}
which are not captured in the present calculation.

The gap opening at the Dirac point and the zone corners
(mini-Dirac points) is all due to the absence of the inversion symmetry
in $V_{\rm hBN}$ which was argued in the previous section.
Generally, the coexistence of the spatial inversion symmetry
and the time reversal symmetry requires vanishing of
the Berry curvature at any nondegenerate points in the energy band, 
\cite{haldane2004berry, fu2007topological}
and this guarantees the robustness of band touching points
in two-dimensional systems. \cite{koshino2013electronic}
Therefore the original Dirac point at $K$ in intrinsic graphene 
is never gapped without breaking either the time-reversal
or the inversion symmetry.
If we have a band touching point
under the time-reversal and the inversion symmetries,
it requires the existence of another band touching point 
somewhere in the same energy band.
This is because if we only have a single Dirac point in the band,
the integrated Berry curvature 
over the superlattice Brillouin zone except for that Dirac point
becomes $\pm\pi$ (from the only Dirac point), and never vanishes.
\footnote[2]{In the twisted bilayer graphene, in contrast,
the lowest energy band has two Dirac points at zero energy 
(from two $K$ points of the top and bottom layer)
so that we can have an energy gap between the first and the second bands
even though the system is inversion symmetric.
}
In the present effective Hamiltonian,
when $V_{\rm hBN}$ is modified by hand
so as to have the inversion symmetry (e.g., $\psi$ is set to $\pi/3$),
we actually see that the adjacent bands touch at either of $X$, $Y$
or $\Gamma$ and all the energy bands are connected. \cite{wallbank2013generic}

The electron-hole asymmetric splitting at the zone corners
can be explained by  the matrix elements in the effective model.
The electronic states at zone corner $X(Y)$ 
are originally from three $k$ points
on the equi-energy surface of the intrinsic Dirac cone
and they are mixed by the effective potential $V_{\rm hBN}$,
as shown by dashed arrows in Fig.\ \ref{fig_unit_cell}(b).
For example, the matrix element between two $X$ points
(denoted as $X_1, X_2$)
connected by $\Vec{G}_2^{\rm M}$ in
Fig.\ \ref{fig_unit_cell}(b)
are obtained by 
\begin{equation}
\langle X_2 |
\,\,
V_1 e^{i\xi \psi}
\begin{pmatrix}
1 & \omega^{\xi}
\\
\omega^{\xi} & \omega^{-\xi}
\end{pmatrix}
| X_1 \rangle,
\end{equation}
where $|X_1 \rangle$ and $|X_2 \rangle$
are Dirac spinors corresponding to the $k$-points,
and the matrix in the middle comes from
the term having $e^{i\xi\Vec{G}_2^{\rm M}\cdot \Vec{r}}$
in Eq.\ (\ref{eq_VhBN}).
The matrix elements connecting the triplets
are shown to be all identical,
and their amplitude determines the energy scale of the band splitting.
In $\theta =0^\circ$, the absolute value of the matrix elements
in units of $V_1$ are shown to be 3/2 and 2 for $X$ and $Y$
on the hole side, respectively, 
and they are actually larger than those for the electron side,
1/2 and 0 for $X$ and $Y$, respectively.

\begin{figure*}
\begin{center}
\leavevmode\includegraphics[width=0.9\hsize]{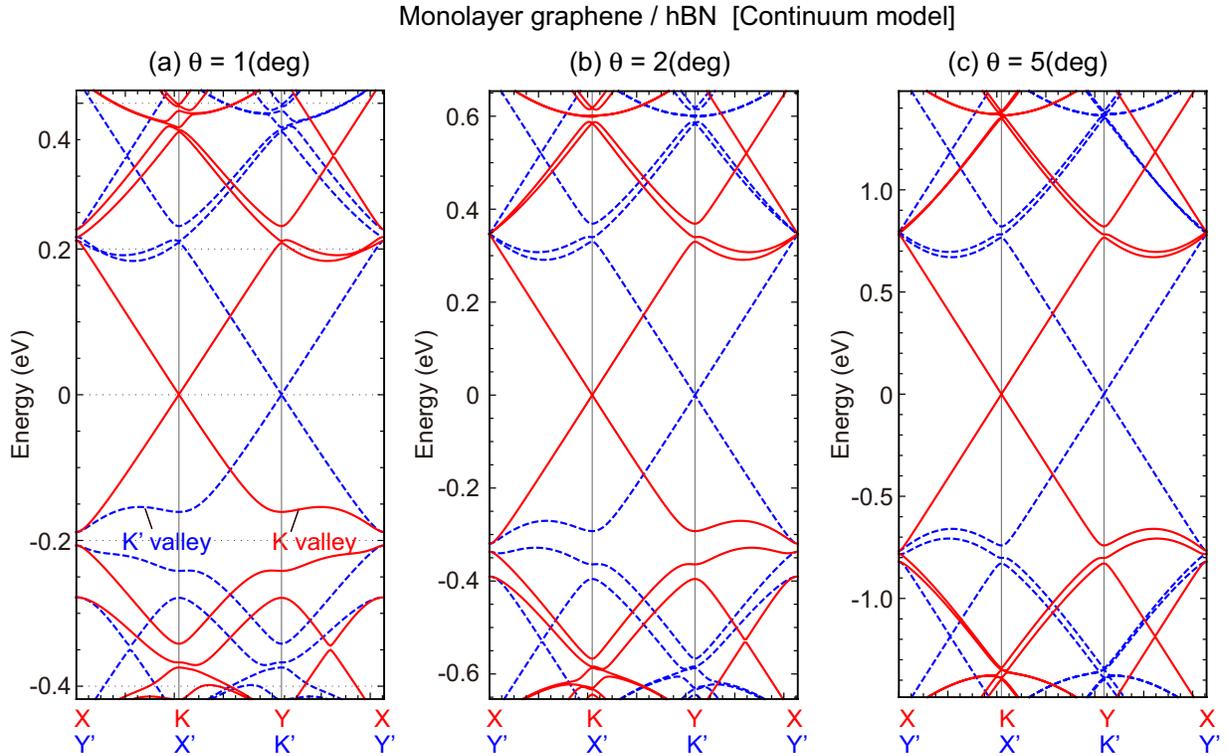}
\end{center}
\caption{
Band structures
of a monolayer graphene / hBN system 
with (a) $\theta=1^\circ$, (b) $2^\circ$ and
(c) $5^\circ$,
calculated by the effective continuum model.
}
\label{fig_band_mono_angle}
\end{figure*}

The continuum model can be easily extended to 
other twist angles, which are generally hard to treat 
in the tight-binding model due to the lattice incommensurability.
Figure \ref{fig_band_mono_angle} plots the band structures
of monolayer graphene + hBN with (a) $\theta = 1^\circ$,
(b) $2^\circ$ and (c) $5^\circ$,
calculated by the continuum model.
We see that the band structures all look similar,
while the energy scale expands in increasing $\theta$,
according to the increase of the 
characteristic scale $2\pi \hbar v/L^{\rm M}$.
At the same time, 
the band splitting, which is of the order of $u_0$,
becomes relatively small compared to the band width.

\begin{figure*}
\begin{center}
%\leavevmode\includegraphics[width=0.9\hsize]{fig_band_bi.eps}
\leavevmode\includegraphics[width=0.9\hsize]{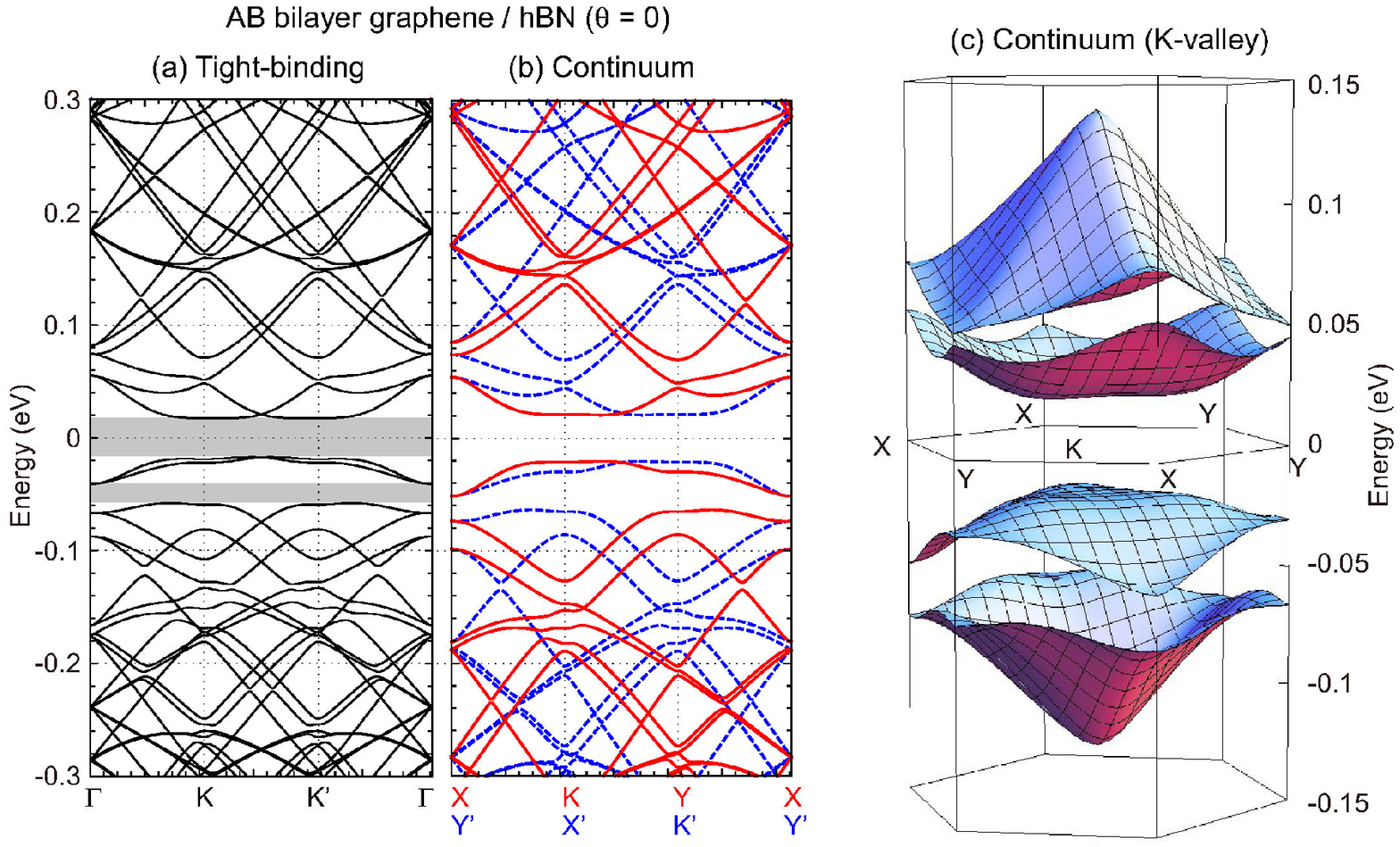}
\end{center}
\caption{Plots similar to Fig.\ \ref{fig_band_mono},
calculated for $AB$-bilayer graphene / hBN system with $\theta=0^\circ$.
}
\label{fig_band_bi}
\end{figure*}

Figure \ref{fig_band_bi} plots the band structure
of $AB$-stacked bilayer graphene on a hBN system stacked at $\theta=0^\circ$.
We see the good agreement between the tight-binding model
and the continuum model.
Unlike a monolayer graphene-hBN system,
we observe a relatively large spectral gap about $40\,\mathrm{meV}$
at zero energy, which is accompanied by flat band edges.
This is actually due to the 
interlayer potential difference in bilayer graphene,
which is caused by $V_0$ terms in $V_{\rm hBN}$ for the layer 1.
The width of the central gap should also depend on the gate electric field
and other electrostatic environments,
which contribute to the interlayer potential asymmetry.
We also see a band gap between the first band and the second band,
and it is larger on the hole side than on the electron side
similar to the monolayer graphene-hBN system.
The recent experiment observed consistent features
where a stronger resistance peak appeared
on the hole side. \cite{dean2013hofstadter}

%The width of the subbands is much narrower than
%monolayer's case
%because of the quadratic dispersion in the intrinsic AB bilayer
%graphene.

%\textcolor{red}{
%The presence of the gap
%is consistent with the strong insulating signal
%at charge neutral point in a recent experiment.\cite{dean2013hofstadter}
%Due to the lift of inversion symmetry by $V_0$ terms,
%}

\section{Spectrum in magnetic field}
\label{spectrum_in_magnetic_field}

We calculate the energy spectrum
of monolayer graphene on a hBN system
under a uniform perpendicular magnetic field.
Here we use the tight-binding 
lattice Hamiltonian \cite{moon2012energy}
with a Peierls phase
\begin{eqnarray}
\phi_{ij} = -\frac{e}{\hbar}\int_j^i \Vec{A}(\Vec{r})\cdot d\Vec{r}
\end{eqnarray}
between sites $i$ and site $j$.
Here $\Vec{A}(\Vec{r}) = (0,Bx,0)$ is the vector potential
giving a uniform magnetic field $B$ perpendicular to the layers.
We take the wave functions of
low-lying Landau levels of monolayer
in $|\vare| \lsim 1.0\,\mathrm{eV}$, 
and compose the Hamiltonian matrix by writing 
$H$ in terms of the reduced basis
\cite{moon2012energy}.
For simplicity, we neglected
spin Zeeman splitting throughout the calculation.

\begin{figure*}
\begin{center}
\leavevmode\includegraphics[width=0.9\hsize]{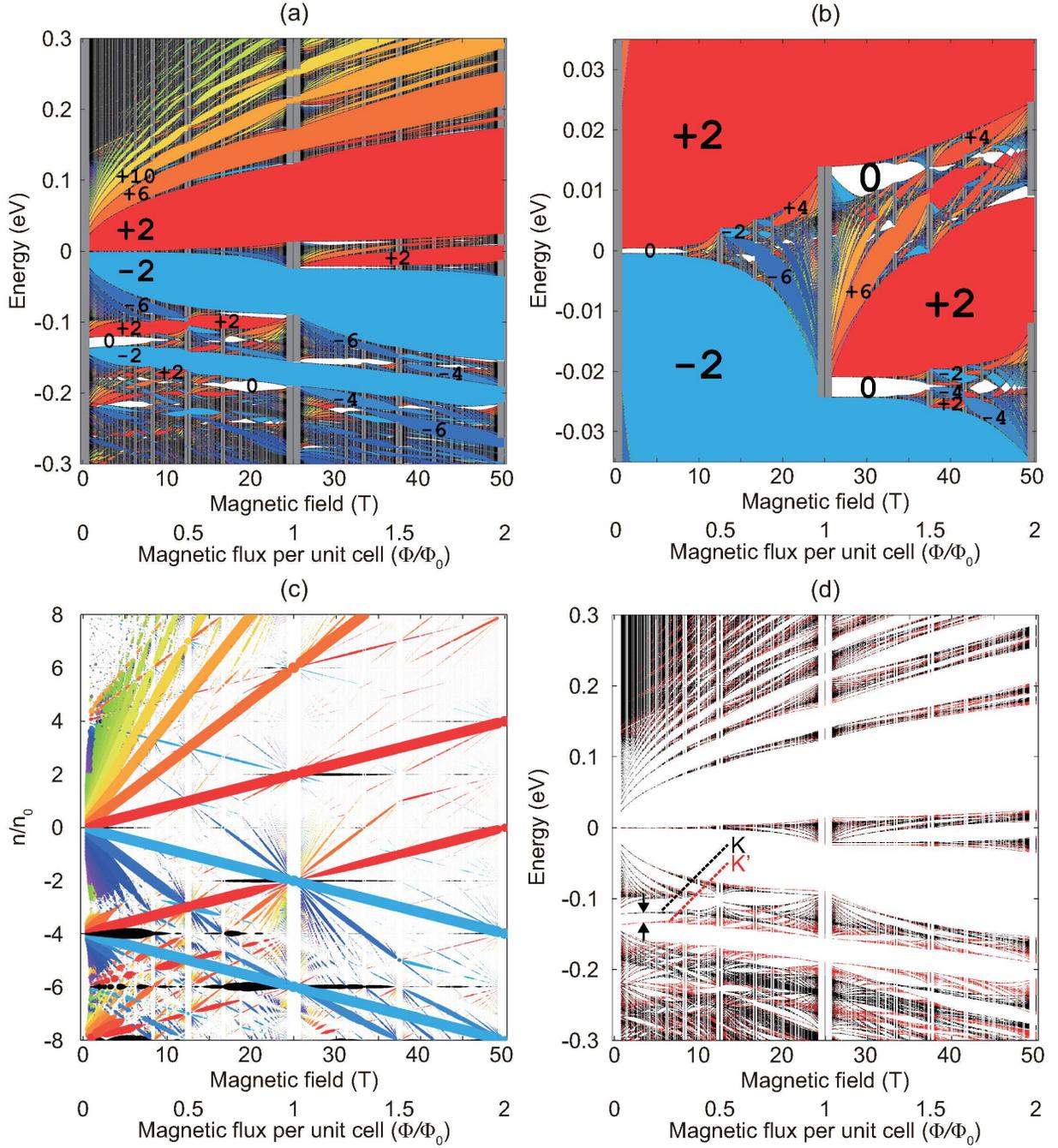}
%\leavevmode\includegraphics[width=0.9\hsize]{Fig_HB_MG_hBN.eps}
\end{center}
\caption{
(Color online)
Energy spectrum
of monolayer graphene on hBN system with $\theta=0^\circ$
as a function of magnetic field strength
in (a) wide and (b) narrow ranges of energy.
In each figure, the quantized values of
Hall conductivity inside energy gaps
are indicated by numbers in units $-e^2/h$
as well as shading filling the gaps.
The Hall conductivity of the gray area
cannot be determined by the present calculation.
(c) Wannier diagram calculated for
the energy spectrum in (a).
Each gap is plotted as
a line of which thickness is proportional to the gap width,
and the color represents the quantized Hall conductivity.
The colormap for the Hall conductivity
is the same as that in (a) and (b),
except the black circle represents
the gap with Hall conductivity 0 in (c).
(d) Energy spectrum originating from monolayer's $K$ region (black)
and $K'$ region (red).
}
\label{fig_hb_mg_hbn}
\end{figure*}

Figure \ref{fig_hb_mg_hbn}(a)
shows the energy spectrum
of graphene monolayer on hBN with $\theta = 0^\circ$,
as a function of magnetic field strength,
with the magnetic flux per a superlattice unit cell $\Phi$
measured in units of $\Phi_0 = h/e$.
The quantized Hall conductivity inside the energy gaps are represented by
numbers in units of $-e^2/h$
as well as shading filling the gaps.
While we concentrate on $\theta = 0^\circ$ in the following discussions,
the spectrum should look similar 
in other twist angles $\theta$ ($< 10^\circ$)
except for the characteristic energy scale,
as naturally expected from the similarity in the 
zero-field band structures in Fig.\ \ref{fig_band_mono_angle}.

On the electron side, the spectrum can be viewed as
the Landau levels of intrinsic monolayer graphene
with the fine structure inside,
while on the hole side, in contrast, the monolayer's Landau levels
are completely reconstructed into the fractal spectrum.
%where the Hall conductivity in varies nonmonotonically 
%as Fermi energy changes.
%\cite{kohmoto1985topological,thouless1982quantized}
This feature coincides with the zero-field band structure,
Fig.\ \ref{fig_band_mono},
in which the hole side is strongly modified by a large gap opening
at the mini-Dirac point.
Figure \ref{fig_hb_mg_hbn}(b) shows the spectrum
near zeroth Landau level.
The width of the modulated Landau level
rapidly grows in the high field region $B > 10\,T$,
where the minigap structure inside the level
becomes significant.

% Valley-degeneracy lift
In a pristine monolayer graphene,
the Landau levels are completely 
 valley ($K$, $K'$) degenerate 
because of the intrinsic inversion symmetry. \cite{koshino2010parity}
As a result, the quantized Hall conductivity
can only have the values of $4m + 2$ ($m \in \mathbb{Z}$),
where the factor 4 is from the spin-valley degeneracy.
In monolayer on a hBN system, the valley degeneracy
is broken by the inversion asymmetric $V_{\rm hBN}$.
In Fig.\ \ref{fig_hb_mg_hbn}(d), we plot the energy spectrum
with different shadings (colors) for $K$ and $K'$ valleys.
We can see that the degeneracy between $K$ and $K'$ levels
is actually lifted, and the levels from different valleys
simply cross each other, since the two valleys are hardly hybridized
by the superlattice potential.
As a consequence of the valley splitting,
we have the Hall conductivity $4m$
outside the standard sequence of monolayer graphene,
as seen in Figs.\ \ref{fig_hb_mg_hbn}(a) and \ref{fig_hb_mg_hbn}(b).

When the Dirac point in graphene 
is gapped by a time-reversal symmetric potential, generally, 
the zeroth Landau level of one valley sticks to the 
top of the gap, while that of the other valley
sticks to the bottom of the gap \cite{koshino2010anomalous}.
Therefore, a larger energy gap in the Dirac spectrum
is always accompanied by the larger valley splitting
in the magnetic field.
In Fig.\ \ref{fig_hb_mg_hbn}(d),
we actually see a large valley splitting of $\sim 20\,\mathrm{meV}$
at the hole-side mini-Dirac point 
(marked by arrows),
which exactly corresponds to the energy gap 
in the zero-field band structure, Fig.\ \ref{fig_band_mono}.
Similarly, the valley splitting of
the zeroth Landau level 
in small $B$ corresponds to a tiny energy gap $2\,\mathrm{meV}$ 
at the central Dirac point.

% Wannier

Figure \ref{fig_hb_mg_hbn}(c) is the Wannier diagram 
\cite{wannier1978result},
which indicates the positions of energy gaps 
in the space of charge density $n$ and magnetic field $B$.
The thickness of each line is proportional to 
the width of the corresponding energy gap.
In the Wannier diagram,
each single energy gap always follows a linear trajectory,
 \cite{kohmoto1985topological,thouless1982quantized}
\begin{eqnarray}
\frac{n}{n_0}=t\frac{\Phi}{\Phi_0}+s,
\end{eqnarray}
where $n$ is the electron density,
$n_0=1/S$ is the electron density per each Bloch band,
and $t$ and $s$ are topologically invariant integers.
The quantized Hall conductivity is 
given by $-t e^2/h$.\cite{kohmoto1985topological,thouless1982quantized}
In the vicinity of the Dirac point at weak-field regime,
we see a conventional Landau fan diagram
where the gap trajectories originate from the charge neutral point
at $B = 0$ (i.e., $s=0$).
In a fractal band regime, on the other hand
we see a different series of trajectories
having nonzero $y$-intercept (i.e., $s\ne0$) at $B = 0$,
which are an evidence of Hofstadter's spectrum. \cite{dean2013hofstadter}
In accordance with the large gap opening 
at the mini-Dirac point of the hole side,
we have strong signals from the mini-Landau fan centered at $n/n_0 = -4$.
At the cross points with the conventional Landau fan
and mini-Landau fan (e.g., $n/n_0=-2$ at $\Phi/\Phi_0=1$)
we have the second generation of the Landau fan 
as a part of the recursive structure.

\begin{figure*}
\begin{center}
\leavevmode\includegraphics[width=0.9\hsize]{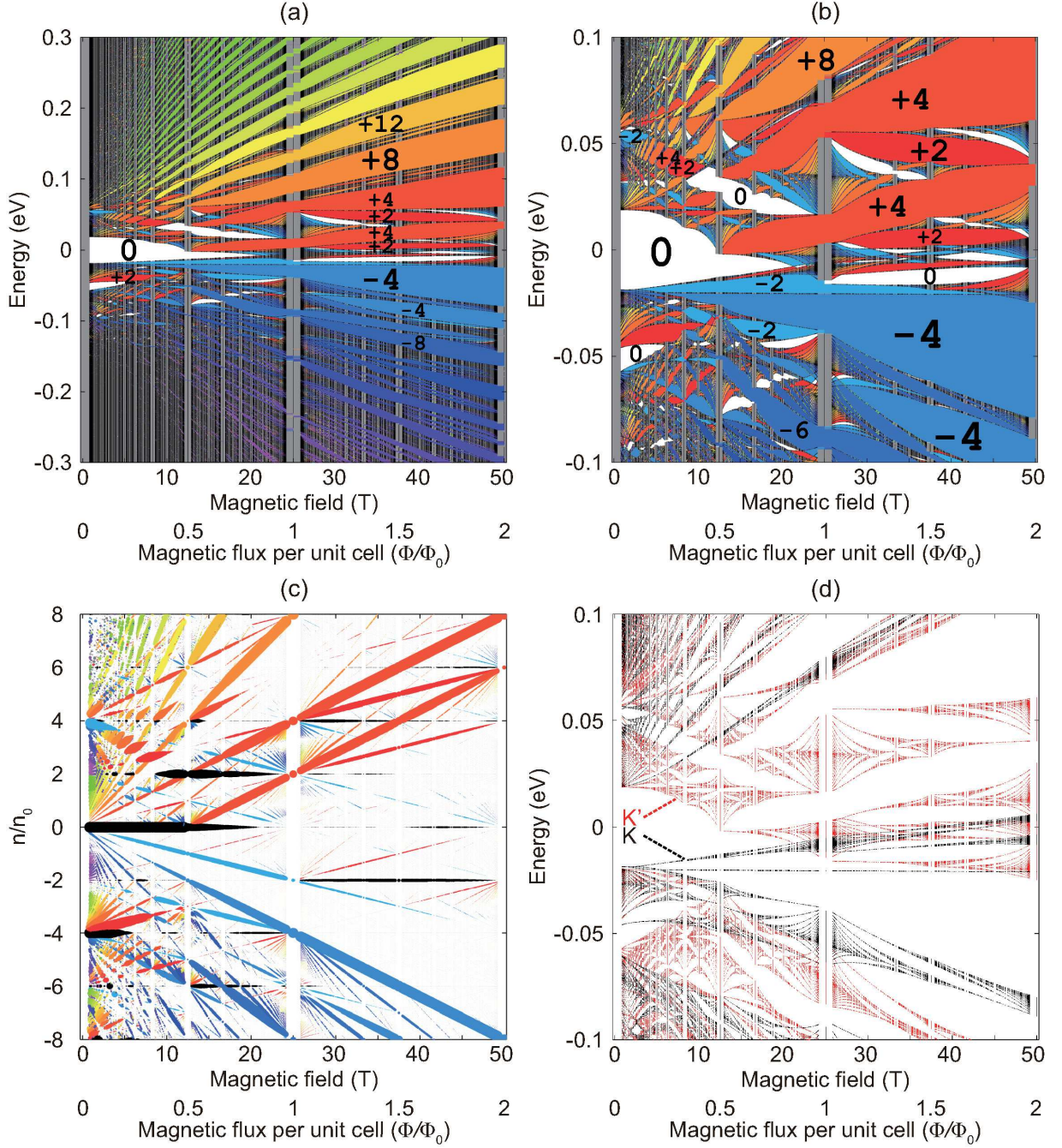}
%\leavevmode\includegraphics[width=0.9\hsize]{fig_HB_AB_hBN.eps}
\end{center}
\caption{
(Color online)
Plots similar to Fig.\ \ref{fig_hb_mg_hbn}
for $AB$-stacked bilayer graphene on a hBN system with $\theta=0^\circ$.
}
\label{fig_hb_ab_hbn}
\end{figure*}

% BG/hBN
Figures \ref{fig_hb_ab_hbn}(a) and \ref{fig_hb_ab_hbn}(b)
show the energy spectrum of
$AB$-stacked bilayer graphene on a hBN system
with $\theta = 0^\circ$
and quantized Hall conductivity in the magnetic field.
The spectrum in the low-energy region $|E| < 0.1\,eV$
exhibits a complicated fractal structure, 
corresponding to the strong modification
in the zero-field band structure.
In the valley separated spectrum, Fig.\ \ref{fig_hb_ab_hbn}(d),
the valley splitting is much greater than
in the monolayer-hBN case.
The spectra of $K$ and $K'$ exhibit completely different 
configurations,
and cannot be regarded as a shifted pair of the same spectrum.
In a bilayer-hBN system, only a single layer out of two graphene layers
feels the effective potential of hBN,
and it severely breaks the inversion symmetry
which swaps the two layers.
In monolayer-hBN system, in contrast,
the inversion symmetry breaking solely comes from 
$\Vec{r}$ dependence of $V_{\rm hBN}(\Vec{r})$,
and the effect is relatively minor.

\begin{figure}
\begin{center}
\leavevmode\includegraphics[width=0.8\hsize]{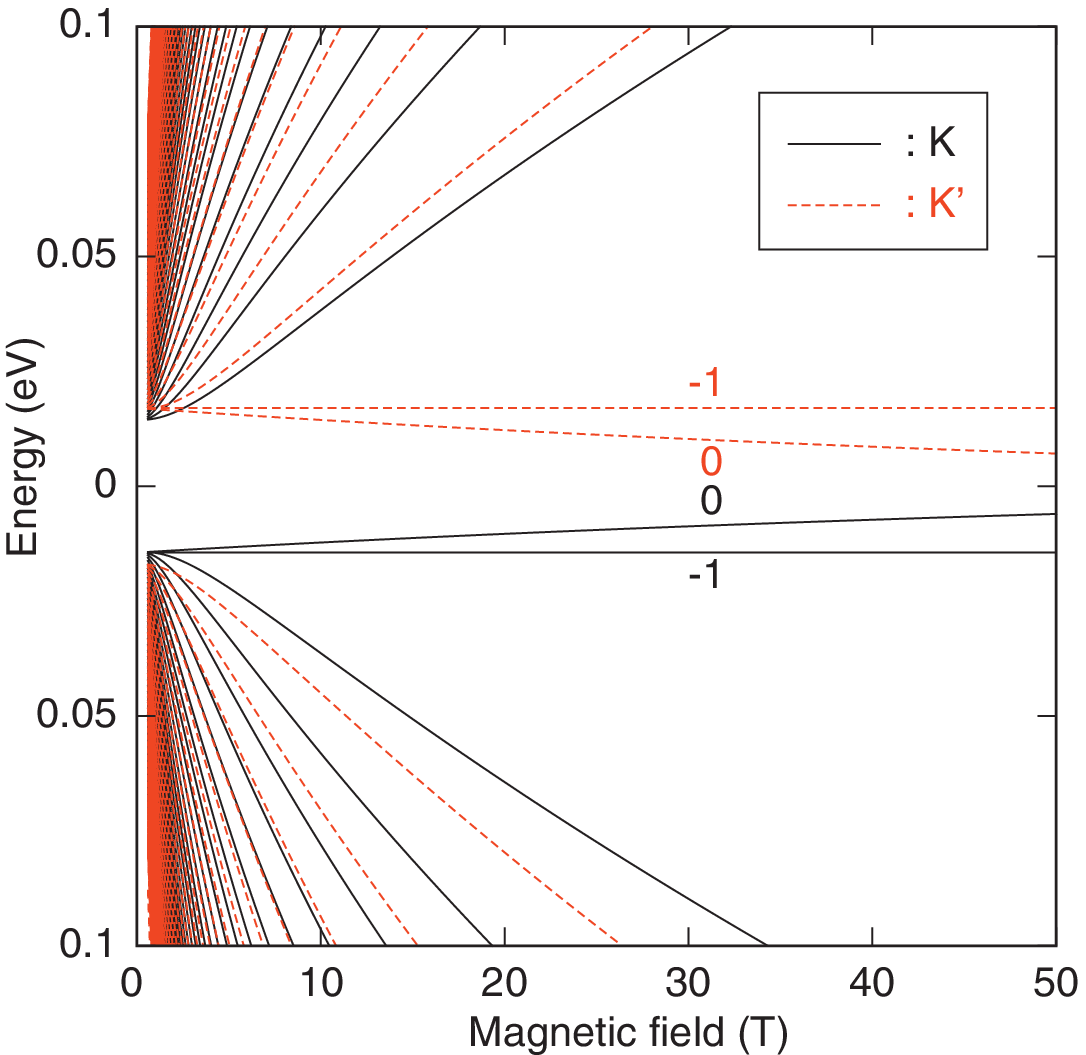}
\end{center}
\caption{
Landau level structure in bilayer graphene
with the interlayer potential asymmetry $\pm V_0/2$.
}
\label{fig_bilayer_LL}
\end{figure}

The complicated level structure in a bilayer-hBN system
can be better understood by comparing the spectrum
to Fig.\ \ref{fig_bilayer_LL}, which plots the Landau levels
of the bilayer graphene with interlayer asymmetric potential
$V_0$ and 0 for the layer 1 and 2, respectively
(the origin of the energy axis is set to the gap center).
This corresponds to the
situation where we neglect all the spatially-modulating 
terms in $V_{\rm hBN}$, leaving only the constant term $V_0$.
The central energy gap
and the valley splitting of $n=-1,0$ levels
roughly coincide with the properties in the original spectrum.
In addition, among the four levels
($n=-1,0$ at $K$ and $K'$) which comprise the zero-energy Landau levels,
only the two levels of $K'$ evolve into a clear fractal spectrum,
and the other two of $K$ remain almost intact.
This is because the wave function of the zero-energy Landau levels
($n=-1,0$) in bilayer graphene are layer-polarized depending on the valley:
$K$ levels are localized on layer 2 
while $K'$ levels are on layer 1.
Since the hBN layer influences layer 1,
the fractal evolution of the spectrum 
is much clearer in $K'$  than $K$.
In $K$ valley, we see that
the $n = -1$ level  remains almost Landau level-like,
while the $n = 0$ level
exhibits a small minigap structure.
This is because the wave function of $n = -1$ 
is almost completely localized on layer 2,
while the state with $n = 0$
has small amplitude on layer 1, which is 
proportional to interlayer asymmetric potential $V_0$.
\cite{koshino2010anomalous,koshino2011landau}

The lift of valley degeneracy
directly affects the quantized value of the Hall conductivity
in Figs.\ \ref{fig_hb_ab_hbn}(a), \ref{fig_hb_ab_hbn}(b)
and \ref{fig_hb_ab_hbn}(c).
In regular bilayer graphene, the Hall conductivity 
can take the values of $4m$ ($m \in \mathbb{Z}$),
where the factor 4 is from the spin-valley degeneracy.
In moir\'{e} system $4m+2$ can also appear 
due to the valley splitting. \cite{dean2013hofstadter}

It should be noted that the Hofstadter butterfly in this work arises 
from the competition between the long-period moir\'{e} superlattice potential 
(of the order of $1-100\, \mathrm{nm}$) and magnetic field. 
On the other hand, there is another rich spectral structure 
which comes from the competition between the atomic lattice period of 
constituent layers (order of $0.1\, \mathrm{nm}$) and magnetic field.\cite{PhysRevB.88.125426,pedersen2013hofstadter}
Considering the condition for the fractal spectrum, $Ba^2/(h/e) \sim 1$, 
the latter effect becomes conspicuous in a relatively high magnetic field range.
In the present calculation, both interference effects are fully taken into account 
in the tight-binding Hamiltonian, while the effect of the atomic periodicity 
is almost negligible in the magnetic range of $0-50\,\mathrm{T}$ considered here.

\section{Conclusion}
\label{conclusion}

We calculated the band structures of
moir\'{e} systems composed of
monolayer and bilayer graphene on the hBN layer.
We developed an effective continuum theory
in the framework of a tight-binding method
and analytically investigated several
characteristic properties in the band structure.
We showed that the inversion-asymmetric term generally opens
an energy gap both at the intrinsic Dirac point
and the mini-Dirac point,
and the gap width exhibits a strong electron-hole asymmetry.
We investigated the energy spectrum and quantum Hall effect 
of graphene-hBN systems in uniform magnetic field,
and demonstrated the evolution 
of the fractal spectrum as a function of the magnetic field.
The lack of the inversion symmetry is responsible for
the breaking of the valley degree of freedom.
The valley splitting is more significant
in bilayer graphene on hBN than in monolayer graphene on hBN
because of the stronger inversion-symmetry breaking in a bilayer-hBN system.

\section*{ACKNOWLEDGEMENTS}
P. M. was supported by New York University Shanghai
Start-up Funds, and appreciate the support from
East China Normal University
for providing research facilities.

P.M. was supported by New York University Shanghai (research funds)
and East China Normal University (research facilities).
M.K. was funded by JSPS Grants-in-Aid for Scientific Research
[Grants No. 24740193 and No. 25107005].

%%%%%%%%%%%%%%%%%%%%

\appendix

\section{Interlayer coupling in nonrotated bilayer of honeycomb lattices}
\label{sec_app_a}

\begin{figure}
\begin{center}
\leavevmode\includegraphics[width=0.4\hsize]{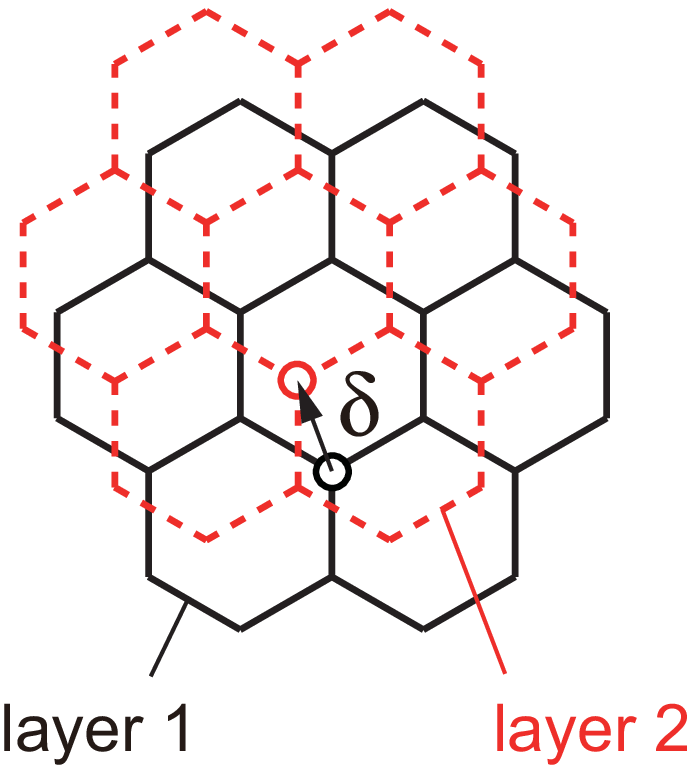}
\end{center}
\caption{
Nonrotated, shifted bilayer of tight-binding honeycomb lattices
with the same lattice constant.
}
\label{fig_shifted_bilayer}
\end{figure}

Here we derive the interlayer coupling Hamiltonian
for a nonrotated, shifted bilayer of tight-binding honeycomb lattices.
We assume the two layers are identical honeycomb lattices
with the same lattice constant,
and they are arranged in parallel fashion
with a constant in-plane displacement $\GVec{\delta}$
and interlayer spacing $d$,
as illustrated in Fig.\ \ref{fig_shifted_bilayer}.
The unit cell includes $A_l$ and $B_l$ for the layer $l=1,2$.
We assume that the transfer integral between any two sites
is given by Eq.\ (\ref{eq_transfer_integral}).

We define the Bloch wave basis of a single layer as
\begin{eqnarray}
&& |\Vec{k},X_l\rangle = 
\frac{1}{\sqrt{N}}\sum_{\Vec{R}_{X_l}} e^{i\Vec{k}\cdot\Vec{R}_{X_l}}
|\Vec{R}_{X_l}\rangle,
%\nonumber \\
%&& |\Vec{k},B_l\rangle = 
%\frac{1}{\sqrt{N}}\sum_{\Vec{R}_{B_l}} e^{i\Vec{k}\cdot\Vec{R}_{B_l}}
%|\Vec{R}_{B_l}\rangle,
\label{eq_bloch_base}
\end{eqnarray}
where $\Vec{k}$ is the Bloch wave vector,
$X = A,B$ is the sublattice index, $l= 1, 2$ is the layer index,
and $N$ is the number of monolayer graphene unit cells
(containing a single pair of $A$ and $B$ sites) in the whole system.
The interlayer matrix element is then written as
\begin{eqnarray}
&& U_{A_2A_1}(\Vec{k},\GVec{\delta}) 
\equiv \langle \Vec{k},A_2| H |\Vec{k},A_1\rangle
= u(\Vec{k},\GVec{\delta}),
\nonumber\\
&& U_{B_2B_1}(\Vec{k},\GVec{\delta}) 
\equiv \langle \Vec{k},B_2| H |\Vec{k},B_1\rangle
= u(\Vec{k},\GVec{\delta}),
\nonumber\\
&& U_{B_2A_1}(\Vec{k},\GVec{\delta}) 
\equiv \langle \Vec{k},B_2| H |\Vec{k},A_1\rangle
= u(\Vec{k},\GVec{\delta} - \GVec{\tau}_1),
\nonumber\\
&& U_{A_2B_1}(\Vec{k},\GVec{\delta}) 
\equiv \langle \Vec{k},A_2| H |\Vec{k},B_1\rangle
= u(\Vec{k},\GVec{\delta} + \GVec{\tau}_1),
\label{eq_interlayer_U}
\end{eqnarray}
where
\begin{eqnarray}
 u(\Vec{k},\GVec{\delta}) = 
\sum_{n_1,n_2}
- t(n_1 \Vec{a}_1 + n_2 \Vec{a}_2 + \Vec{d}_z + \GVec{\delta})
\nonumber\\
\hspace{20mm}
\times
\exp\left[-i\Vec{k}\cdot(n_1 \Vec{a}_1 + n_2 \Vec{a}_2 + \GVec{\delta})
\right],
\label{eq_ukq_1}
\end{eqnarray}
and $\GVec{\tau}_1 = (-\Vec{a}_1+2\Vec{a}_2)/3$ is a vector
connecting the nearest $A$ and $B$ sublattices,
and $\Vec{d}_z = d_{\rm G-hBN}\,\Vec{e}_z$ is the 
perpendicular displacement between graphene and hBN.

The function $u(\Vec{k},\GVec{\delta})$ is obviously
periodic in $\GVec{\delta}$ with periods $\Vec{a}_1$ and $\Vec{a}_2$,
and it is then Fourier transformed as
\begin{eqnarray}
 u(\Vec{k},\GVec{\delta}) = 
\sum_{m_1,m_2} \tilde{t}(m_1\Vec{a}^*_1+m_2\Vec{a}^*_2+\Vec{k})
\nonumber\\
\hspace{20mm}
\times
\exp[
i(m_1\Vec{a}^*_1+m_2\Vec{a}^*_2)\cdot \GVec{\delta}
],
\label{eq_ukq_2}
\end{eqnarray}
where $\tilde{t}(\Vec{q})$ is the in-plane Fourier transform
of $t(\Vec{R})$ defined by
\begin{eqnarray}
 \tilde{t}(\Vec{q}) = 
\frac{1}{S} \int t(\Vec{R}+ \Vec{d}_z) 
e^{-i \Vec{q}\cdot \Vec{R}} d\Vec{R},
\end{eqnarray}
with $S = |\Vec{a}_1\times\Vec{a}_2|$, and the integral in $\Vec{R}$
is taken over an infinite two-dimensional space.
In the present tight-binding model,
$t(\Vec{R})$ exponentially decays in $R \, \gsim \, r_0$, so that
the Fourier transform $\tilde{t}(\Vec{q})$ decays in $q \, \gsim \, 1/r_0$.
In Eq.\ (\ref{eq_ukq_2}), therefore, we only need to take
a few Fourier components within 
$|m_1\Vec{a}^*_1+m_2\Vec{a}^*_2+\Vec{k}| \,\lsim\, O(1/r_0)$.

In the following 
we only consider the electronic states near $\Vec{K}_\xi$ point,
and then we can approximate $u(\Vec{k},\GVec{\delta})$
with $u(\Vec{K}_\xi,\GVec{\delta})$.
Equation (\ref{eq_ukq_2}) then becomes
\begin{eqnarray}
u(\Vec{K}_\xi,\GVec{\delta})  \approx 
u_0 \left[
1 + e^{i \xi \Vec{a}^*_1 \cdot \mbox{\boldmath \scriptsize $\delta$}}
+ e^{i \xi (\Vec{a}^*_1 + \Vec{a}^*_2)\cdot  \mbox{\boldmath \scriptsize $\delta$}}
\right],
\end{eqnarray}
with
\begin{eqnarray}
 u_0 = \tilde{t}(\Vec{K}_\xi),
\end{eqnarray}
which gives Eq.\ (\ref{eq_u0}).
In the present tight-binding parameter, we have $u_0 \approx 0.152\,\mathrm{eV}$.
The second largest Fourier component
is $\tilde{t}(2\Vec{K}_\xi) \approx 0.0025\, \mathrm{eV}$
and is safely neglected.

Finally, Eq.\ (\ref{eq_interlayer_U})
becomes
\begin{eqnarray}
&& U_{A_2A_1}
=  U_{B_2B_1}
= u_0 \left[
1 + e^{i \xi \Vec{a}^*_1 \cdot \mbox{\boldmath \scriptsize $\delta$}}
+ e^{i \xi (\Vec{a}^*_1 + \Vec{a}^*_2)\cdot  \mbox{\boldmath \scriptsize $\delta$}}
\right],
\nonumber\\
&& U_{B_2A_1}
= u_0 \left[
1 + 
\omega^{\xi}
e^{i \xi \Vec{a}^*_1 \cdot \mbox{\boldmath \scriptsize $\delta$}}
+ 
\omega^{-\xi}
e^{i \xi (\Vec{a}^*_1 + \Vec{a}^*_2)\cdot  \mbox{\boldmath \scriptsize $\delta$}}
\right],
\nonumber\\
&& U_{A_2B_1}
= u_0 \left[
1 + 
\omega^{-\xi}
e^{i \xi \Vec{a}^*_1 \cdot \mbox{\boldmath \scriptsize $\delta$}}
+ 
\omega^{\xi}
e^{i \xi (\Vec{a}^*_1 + \Vec{a}^*_2)\cdot  \mbox{\boldmath \scriptsize $\delta$}}
\right].
\label{eq_interlayer_U2}
\end{eqnarray}

In the moire system,
$\GVec{\delta}$ is not constant but slowly depends on the position
$\Vec{r}$.
By replacing $\GVec{\delta}$ in Eq.\ (\ref{eq_interlayer_U2})
with $\GVec{\delta}(\Vec{r})$ in Eq.\ (\ref{eq_delta_of_r}),
we obtain the interlayer Hamiltonian $U$ for the moir\'{e} system,
Eq.\ (\ref{eq_U}).
Here we used the relation 
$\Vec{a}^*_i\cdot\GVec{\delta} = \Vec{G}^{\rm M}_i\cdot \Vec{r}$.

%The amplitude $u_0$ can be derived from $u(\Vec{K}_\xi,0)$ as
%\begin{eqnarray}
%u_0 = \frac{1}{3}t(0,d) - t(a,d) + 2t(\sqrt{3}a,d) -\cdots,
%\end{eqnarray}
%where $t(x,z) = t(x \Vec{e}_x + z \Vec{e}_z)$.

\section{AB bilayer and BA bilayer with hBN}
\label{subsection:ab_bilayer_and_ba_bilayer_with_hBN}

\begin{figure*}
\begin{center}
%\leavevmode\includegraphics[width=0.9\hsize]{Fig_band_bi_diff.eps}
\leavevmode\includegraphics[width=0.9\hsize]{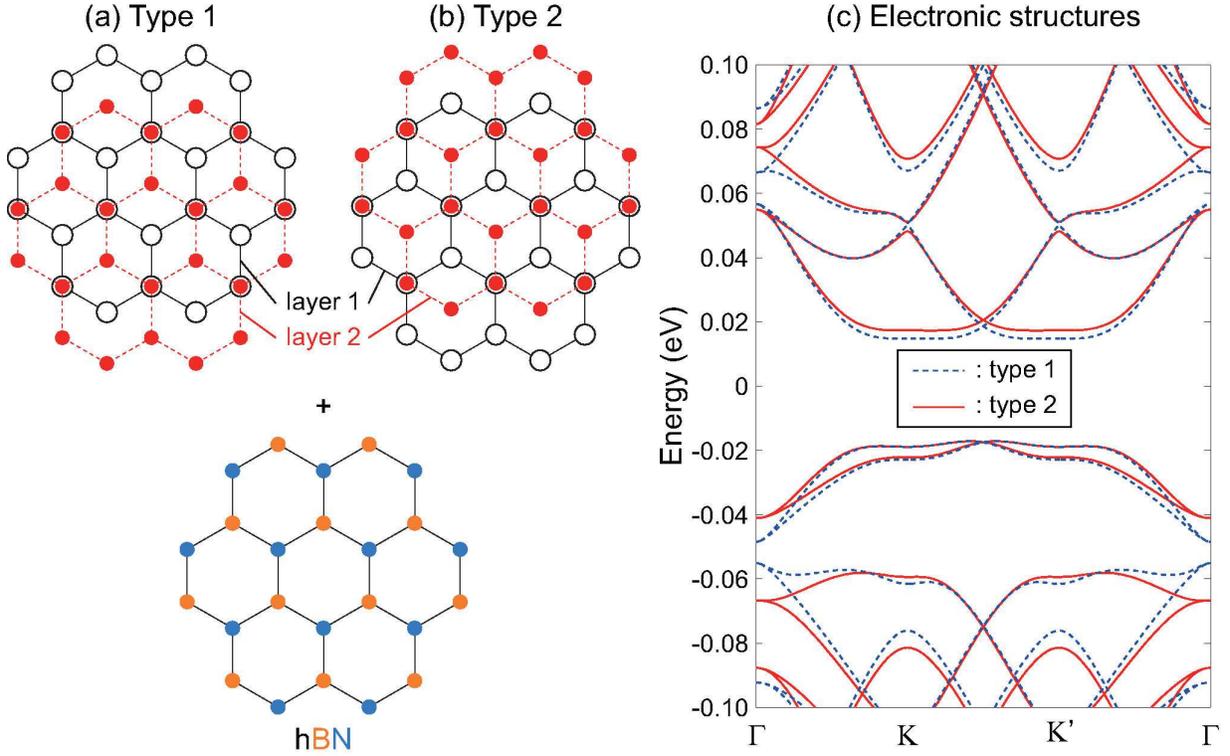}
\end{center}
\caption{
(Color online)
(a) and (b): two different configurations of $AB$-stacked bilayer graphene
on a hBN layer. (c) Electronic structures of the configurations
(a) (dotted blue) and (b) (solid red).
}
\label{fig_band_bi_diff}
\end{figure*}

For $AB$-stacked bilayer + hBN system,
we have two different stacking geometries
illustrated by Figs.\ \ref{fig_band_bi_diff}(a)
and \ref{fig_band_bi_diff}(b),
which we call type 1 and type 2, respectively.
The effective Hamiltonian for type 1 is given by
Eq.\ (\ref{eq_H_BLG-hBN}), and for type 2 is 
\begin{eqnarray}
 {\cal H}^{\rm (type 2)}_{\rm BLG-hBN} = 
\begin{pmatrix}
H_{\rm G}  + V_{\rm hBN} & U_{\rm BLG} \\
U_{\rm BLG}^\dagger & H_{\rm G}
\end{pmatrix},
\label{eq_H_BLG-hBN_type2}
\end{eqnarray}
which is actually distinct from 
Eq.\ (\ref{eq_H_BLG-hBN})
in that the off-diagonal blocks are interchanged.
Figure \ref{fig_band_bi_diff}(c)
compares the energy spectra of type 1 and type 2,
calculated by the tight-binding model and the effective continuum model.
There are small but finite differences in the band structures,
especially at the BZ boundary.

\bibliography{Qiqqa2BibTexExport}

\end{document}